\documentclass[twocolumn,showpacs,preprintnumbers]{revtex4}
\usepackage{amssymb}
\usepackage{amsfonts}
\usepackage{amsmath}
\usepackage{graphicx}
\usepackage{dcolumn}
\usepackage{bm}

\input{tcilatex}

\begin{document}

\title{Quantum theory as a tool for the description of simple psychological
phenomena}
\author{E. D. Vol}
\email{vol@ilt.kharkov.ua}
\affiliation{B. Verkin Institute for Low Temperature Physics and Engineering of the
National Academy of Sciences of Ukraine 47, Lenin Ave., Kharkov 61103,
Ukraine.}
\date{\today }

\begin{abstract}
We propose the consistent statistical approach for the quantitative
description of simple psychological phenomena using the methods of quantum
theory of open systems (QTOS).Taking as the starting point the K. Lewin's
psychological field theory we show that basic concepts of this theory can be
naturally represented \ in the language of QTOS. In particular provided that
all stimuli acting on psychological system (that is individual or group of
interest) are known one can associate with these stimuli corresponding
operators and after that to write down the equation for evolution of density
matrix of the relevant open system which allows one to find probabilities of
all possible behavior alternatives. Using the method proposed we consider in
detail simple model describing such interesting psychological phenomena as
cognitive dissonance and the impact of competition among group members on
its unity.
\end{abstract}

\maketitle

\section{Introduction}

It is a common opinion that in spite of various theories and impressive
concrete results confirmed by numerous experiments modern psychology is
still far from status of exact science such as for example theoretical
physics. The main difference between these sciences is that in theoretical
physics we have well defined concepts and general principles such as for
example action and principle of least action in mechanics or entropy and
second law in thermodynamics which let one to obtain all results of the
theory by successive deductive procedure from general principles. On the
other hand in psychology throughout its history many attempts were taken to
bring together its concepts and facts into integral system that would allow
one to describe and explain known psychological phenomena and possibly to
predict some new effects. In the present paper we start from one of such
theories namely psychological field theory (PFT) of Kurt Lewin and make an
attempt to represent basic concepts of PFT in the language of quantum theory
of open systems (QTOS). This seemingly formal representation has however
indisputable advantage since allow one to use well developed mathematical
methods of QTOS to analyze a variety of psychological effects and
situations. In particular we show that such widely known phenomenon in
psychology as cognitive dissonance(CD) can be consistently considered in the
framework of the method proposed. In addition we demonstrate that this
method can be used to describe group dynamics processes as well as
individual behavior. The rest of the paper is organized as follows. In the
Sect2 we give a brief account of information from PFT and QTOS which is
necessary for understanding of the paper. In the Sect.3 which is major in
the paper we realize representation of such basic concepts of the PFT as the
life space,regions, locomotions and so on in the language of QTOS and
introduce essential concept of density matrix of psychological system (PS)
which let one to find probabilities of all possible behavior alternatives
and formulate the mathematical method for description of evolution this
matrix. In the Sect.4 we demonstrate the effectiveness of our approach in
the framework of simple but useful model for describing several
phychological phenomena such as cognitive dissonance and so forth. Now let
us turn to a detailed account of the paper.

\section{Preparatory Information}

The PFT was created by K.Lewin in the middle of XX century and stated him in
various books and papers (see for example \cite{1a}, \cite{2b}). The
comprehensive review of this theory a reader can find in the textbook of
Hall and Lindzey \cite{3b}. Let us outline the main points of this theory in
the form which is sufficient for understanding of the paper. The initial
concept of the PFT is the concept of the life space by which K.Lewin meant a
set of psychological facts (i.e. both personal incentives and external
impacts connected with situation) acting at the moment on PS and determining
its further evolution. For the behavior description K. Lewin proposed
general but somewhat abstract formula: $B=F(S,P)$, where $B$ means behavior, 
$S$-situation and $P$-person. Note in this connection that the collection of
all variables describing both $S$ and$P$ exactly constitute the life space
of PS. In addition the life space of PS can be divided on separate regions
each of which corresponds to a single psychological fact(for example it may
be separate regions connected with job,family,game and so on)We assume that
different regions are mutually disjoint.Another essential concept in PFT is
a concept of a stimulus.Simple stimuli can be introduced in PFt as certain
psychological forces which act on PS and stimulate it either to occupy
definite region or to avoid it. Depending on the nature of a stimulus we can
attribute to each region definite sign (plus if a stimulus is attractive and
minus in opposite case). The value of a stimulus \ that is the tendency of
PS to occupy given region or to avoid it determines the valence of
corresponding region.It should be noted that in addition to simple stimuli
there are also complex stimuli acting on PS of interest and composed from
several simple ones.Another concept in PFT is locomotion which implies the
transition (but only psychological not physical!) from certain region to
another one. Note that K.Lewin proposed also simple graphic method which
allows one to represent completely the current state of PS within the
framework of these basic concepts. According to K. Lewin such representation
is sufficient to explain actual behavior of PS in future.

However it should be noted that due to extraordinary complexity of the
psychological phenomena modern psychology prefers not to talk about
deterministic laws of human behavior but rather only on its statistical
description.(see for example \cite{4b}). Therefore in the rest of the paper
we assume the task of behavior description is solved if we can indicate the
distribution function which determines probabilities of all behavior
alternatives or by other words the probabilities of finding PS in arbitrary
region of its life space..

Since the main goal of our paper is the representation of basic concepts of
the PFT in the language of quantum theory let us remind the necessary
information from QTOS. First of all as long as evolution of quantum open
system is nonunitary its state should be specified with the help of density
matrix (but not wave function). The main result from QTOS that we need in
this paper is the Lindblad equation which describes evolution of density
matrix in the case of Markov open quantum system.This equation has the next
general form:

\begin{equation}
\frac{\partial \widehat{\rho }}{\partial t}=-\frac{i}{\hbar }\left[ \widehat{%
H},\widehat{\rho }\right] +\sum\limits_{i}\left[ \widehat{R}_{i}\widehat{%
\rho },\widehat{R^{+}}\right] +\ \ h.c,  \label{k1}
\end{equation}

In Eq. (\ref{k1}) $\widehat{H}$ \ is Hermitian operator ("hamiltonian" of
open system) and $\widehat{R_{i}}$ are non-Hermitian operators that specify
all connections of open system of interest with its environment.If the
initial state of the system $\rho (0)$ is known Eq. (\ref{k1}) allows one to
find the behavior( i.e. its state at any time t and thus to determine
average values of all observables relating to open system). If we are
interested only in stationary states of open system we must equate r.h.s. of
the Eq. (\ref{k1}) to zero and find stationary density matrix from this
equation. Before moving on we want to explain \ one essential point namely :
why quantum in nature Eq. (\ref{k1}) can be used to describe \ behavior of
classical systems? The answer is that density matrix of quantum system
represents its classical correlations as well as quantum.But information
about classical correlations is contained in diagonal elements of density
matrix while the information about quantum correlations in nondiagonal ones.

Thus in the situation when we are able to write closed equations including
only a set of diagonal elements of density matrix \ and to solve them, we
actually obtain classical distribution function and required description of
classical analog of corresponding quantum system. As the author has shown in 
\cite{5b} such case holds for example in the case when all operators $R_{i}$
in Eq. (\ref{k1}) have monomial form namely $R_{i}\backsim \left(
a^{+}\right) ^{k_{i}}\left( a\right) ^{l_{i}}$ \ (where operators $a$ and $%
a^{+}$ are bose operators with standard commutation rules: $\left[ \widehat{a%
},\widehat{a^{+}}\right] =1$). In this case Eq. (\ref{k1}) can be reduced to
closed system of equations for diagonal elements of $\rho \left( \widehat{N}%
\right) $ (where $\widehat{N}=\widehat{a^{+}}\widehat{a}$ is number
operator). These arguments from \cite{5b} can be extended also to the case \
when operators $R_{i}$ are represented as monomial forms of some fermi
operators $f_{_{j}}$ and $\widehat{f_{k}^{+}}$ (in the case of several
degrees of freedom ) and corresponding number operators $\widehat{N_{i}}=%
\widehat{f_{i}^{+}}\widehat{f_{i}}$ \ have only two eigenvalues 0 and 1. In
the next part we will demonstrate as formalism of QTOS with the help of such
operators can be used for the statistical description of PS systems in the
language of PFT.

\section{Mapping between PFT and QTOS}

In this part we propose the representation of all basic concepts of PFT in
the language of QTOS.Note, that we will restrict ourselves to considering
only such phenomena when behavior of the PS of interest is determined only
by variables describing a situation while variables associated with a person
play minor role.In addition we will assume that in the process of behavior
there is no restructuring of the life space(that is the number of regions
and all their characteristics remain fixed). Such PS we denote as "simple
system". The generalization of the method proposed that takes into account
also personal variables of PS will be considered elsewhere. Obviously, or PS
under consideration is inside the given region of the life space or outside
it. In accordance with this fact we can attribute to every region occupation
number $n_{i}$ which takes only two values $n_{i}=1$ if PS is in the $i$
region and $n_{i}=0$ in opposite case. Let us introduce operator $\widehat{%
n_{i}}$ acting in two-dimensional linear space whose eigenvalues are 0, 1.
Let us introduce also the pair of operators $\widehat{f_{i}^{+}}=%
\begin{pmatrix}
0 & 1 \\ 
0 & 0%
\end{pmatrix}%
$and $\ \widehat{f_{i}}=%
\begin{pmatrix}
0 & 0 \\ 
1 & 0%
\end{pmatrix}%
$ thus that relationship $\widehat{n_{i}}=\widehat{f_{i}^{+}}\widehat{f_{i}}=%
\begin{pmatrix}
1 & 0 \\ 
0 & 0%
\end{pmatrix}%
$ \ holds.To avoid misunderstanding we want to emphasize that although the
operators $\widehat{f_{i\text{ \ }}}$ and $\widehat{f_{i}^{+}}$ \ satisfy to
relationship :$\widehat{f_{i}^{+}}\widehat{f_{i}}$ \ $+\widehat{f_{i}}%
\widehat{f_{i}^{+}}=1$ nevertheless studied PS systems are of course
classical (not Fermi systems!). Correspodently all regions of the life space
should be considered as distinguishable. Now we establish the mapping
between different stimuli acting on PS of interest and corresponding
operators acting on states of relevant quantum open system. We assume that
every operator $R_{i\text{ }}$entering in the Lindblad equation for relevant
system corresponds \ to certain stimulus(simple or complex) acting on PS of
interest.Besides we suppose that all $R_{i}$ are some monomial functions of
operators $f_{j}^{+}$ and $f_{k}$ (note that index $i$ enumerates different
stimuli acting on a PS).Note also that the number of regions can differ from
number of stimuli. Let us now formulate two main correspondence rules
between acting stimuli and operators $R_{i\text{ }}$. Rule1: If a stimulus $%
i $ is simple then we associate with it either operator $A_{i}=k_{i}$ $%
f_{i}^{+}$ if given stimulus is attractive i.e. stimulates PS to occupy $i$
region or operator $B_{i}=l_{i}$ $f_{i}$ in opposite case. Coefficients $k_{i%
\text{ }}$and $l_{i}$ reflect the value of acting stimulus $i$. Rule2: Let a
stimulus is complex ( that is composed from several simple ones) then in the
case when among simple stimuli $i_{1},i_{2}...$are attractive and stimuli $%
j_{1},j_{2}...$are repulsive with such complex stimulus we associate the
operator $C=k$ $f_{i_{1}}f_{i_{2}}...f_{j_{1}}^{+}f_{j_{2}}^{+}...$Note that
sign of the coefficient $k$ does not affect the final result. In addition
the current state of PS can be represented by \ density matrix of behavior $%
\widehat{\rho \left( t\right) }$ in the linear space of dimension $2^{N}$ \
which is the tensor product $H_{1}\otimes H_{2}\otimes ...H_{r}$ $\left(
1\leq r\leq N\right) $, where $N$ is number of different regions in the life
space). Every space $H_{\alpha }$ $\left( \alpha =1,....N\right) $ is
two-dimensional vector space with basis$\ $states $%
\begin{pmatrix}
1 \\ 
0%
\end{pmatrix}%
_{\alpha }$ and $%
\begin{pmatrix}
0 \\ 
1%
\end{pmatrix}%
_{\alpha }$ connected with $\alpha $ region.In the language of PFT the state$%
\begin{pmatrix}
1 \\ 
0%
\end{pmatrix}%
_{\alpha }$ corresponds to PS which occupies $\alpha $ region of the life
space and the state$%
\begin{pmatrix}
0 \\ 
1%
\end{pmatrix}%
_{\alpha \text{ \ }}$ corresponds to PS which is outside of it. Guided by
these simple rules of correspondence one can easily represent any current
situation with PS as PFT draws it using the rigorous language of QTOS. If we
assume besides that considered PS has no memory then one can try to use for
the description of its evolution the Lindblad equation Eq. (\ref{k1}) for
density matrix of relevant quantum open system.(with $H=0$ ). Operators $%
R_{i}$ in this equation should be chosen of course in correspondence with
the rules stated above. Thus our main \ assumption in the present paper is
the assertion that detail describing of behavior of PS can be realized in
the language of QTOS at least as well as by concepts of PFT.But now we have
in hands powerful mathematical formalism which is sufficient for
quantitative description of various although at present time only simple
phychological phenomena. Since the previous consideration to a considerable
extent was a heuristic now we want with reference to concrete psychological
model to demonstrate its effectiveness. We believe that the question about
applicability of the method can be solved only by careful comparision of
obtained theoretical predictions relating to this and similar models and
results observed in test experiments .

\section{Basic Psychological Model}

In this part of the paper we consider simple but useful model illustrating
in our opinion the effectiveness of the method proposed. We begin by
considering such well known phenomenon in psychology as cognitive dissonance
(CD). According to definition (see e.g. \cite{6b}) cognitive dissonance
emerges when a person has two or several ideas (cognitions) which contradict
each other. The state of CD\ occurs for example in a situation of choice
when it is impossible to estimate for sure pros and cons of different
alternatives. Clearly the behavior in such sutuation will necessarily be
random in its nature. Using the correspondence between PFT and QTOS
specified above it is easy to formulate the situation of CD in the language
of QTOS.Let us consider for simplicity the case when a person has only two
positive alternatives with different incentives (which partially contradict
each other) . We can represent such situation with the help of three
operators $R_{i\text{ }}$acting on density matrix $\widehat{\rho \text{ }}$%
of relevant system describing a behavior of such PS . These operators are: $%
\widehat{R_{1}}=\sqrt{\frac{a}{2}}\widehat{f_{1}^{_{+}}},\widehat{R_{2}}=%
\sqrt{\frac{b}{2}}\widehat{f_{2}^{+}}$ \ and$\widehat{\text{ }R_{3}}=\sqrt{%
\frac{c}{2}}\widehat{f_{1}}\widehat{f_{2}}$ . Coefficients of the operators $%
\widehat{R_{i}}$ characterisize values of correspondent stimuli in
convinient normalization.Now according to our assumption ( since PS of
interest has no memory)its behavior can be described by the Lindblad
equation for diagonal elements of density matrix which can be interpreted as
distribution function for a person to make appropriate choice .

\begin{widetext}
\begin{equation}
\frac{%
\partial \rho _{N_{1},N_{2}}}{\partial t}=a\left[ N_{1}\rho _{\bar{N}%
_{1},N_{2}}-\bar{N}_{1}\rho _{N_{1},N_{2}}\right] +b\left[ N_{2}\rho _{N_{1,}%
\bar{N}_{2}}-\bar{N}_{2}\rho _{N_{1},N_{2}}\right] +c\left[ \bar{N}_{1}\bar{N%
}_{2}\rho _{\bar{N}_{1},\bar{N}_{2}}-N_{1}N_{2}\rho _{N_{1},N_{2}}\right] 
\label{k2}
\end{equation}
\end{widetext} where we introduce convinient notation: $\bar{N}_{i}\equiv
1-N_{i}$ $\left( i=1,2\right) $. We are interesting further only in
stationary solutions of Eq. (\ref{k2}) for which the condition $\frac{%
\partial \rho }{\partial t}=0$ is satisfied. In this case matrix equation
Eq. (\ref{k2}) taking into account the normalization condition$%
\sum\limits_{N_{1}N_{2}}\rho \left( N_{1},N_{2}\right) =1$ can be written in
the form of next four linear equations:

\begin{eqnarray}
-a\rho \left( 0,0\right) -b\rho 0,0)+c\rho \left( 1,1\right) &=&0,
\label{3a} \\
-a\rho \left( 0,1\right) +b\rho \left( 0,0\right) &=&0  \label{3b} \\
a\rho \left( 0,0\right) -b\rho \left( 1,0\right) &=&0  \label{3c} \\
a\rho \left( 0,1\right) +b\rho \left( 1,0\right) -c\rho \left( 1,1\right)
&=&0  \label{3d}
\end{eqnarray}%
We can easily to write down the solution of the system Eq. (\ref{3a})- Eq. (%
\ref{3d}) in explicit form:

\begin{eqnarray}
\rho \left( 0,0\right) &=&\frac{abc}{\Delta _{1}}\text{, }\rho \left(
0,1\right) =\frac{b^{2}c}{\Delta _{1}}\text{,}  \label{k4} \\
\rho \left( 1,0\right) &=&\frac{a^{2}c}{\Delta _{1}}\text{, }\rho \left(
1,1\right) =\frac{ab\left( a+b\right) }{\Delta _{1}}\text{,}  \notag
\end{eqnarray}%
where $\Delta _{1}=ab\left( a+b+c\right) +c\left( a^{2}+b^{2}\right) $. Note
that we considered somewhat more general case then usual cognitive
dissonance. Usually assumed that two competing cognitions incompatable.
Evidently this case realized when fractions $\frac{a}{c}$ and $\frac{b}{c}$
tend to zero. In this case the probabilities of possible outcomes are:

\begin{eqnarray}
\rho \left( 0,0\right) &=&\frac{ab}{\Delta _{2}}\text{, }\rho \left(
0,1\right) =\frac{b^{2}}{\Delta _{2}}\text{,}  \label{k5} \\
\rho \left( 1,0\right) &=&\frac{a^{2}}{\Delta _{2}}\text{, }\rho \left(
1,1\right) =0\text{,}  \notag
\end{eqnarray}%
where $\Delta _{2}=a^{2}+b^{2}+ab$. If in addition we assume that cognitions
1 and 2 have identical attractiveness (the case of Buridan donkey) then the
values of different alternatives are: $\rho \left( 0,0\right) =\rho \left(
0,1\right) =\rho \left( 1,0\right) =\frac{1}{3}$ and $\rho \left( 1,1\right)
=0$. Thus our statistical approach results in that probability for a donkey
to die of hunger is only $\frac{1}{3}$. It is worth to note also that
independently from values of coefficients $a,b,c$ for considered\ PS \ we
have the relation

\begin{equation}
\rho \left( 0,1\right) \cdot \rho \left( 1,0\right) =\rho ^{2}\left(
0,0\right) .  \label{k6}
\end{equation}%
This simple relation which does not depend on parameters of the system can
serve as the useful test to prove or disprove the validity of approach
proposed. Now we want to demonstrate that the method proposed can be also
applied for desribing simple processes of group dynamics. Let us assume we
have the same mathematical model that we used for the description of
cognitive dissonance but now look at it with another point of view. We
believe now that this model could describe the behavior of two members of
formal group which tend to reach specific personal goal or status but \ in
their activity compete with each other .We will consider the state $%
\begin{pmatrix}
1 \\ 
0%
\end{pmatrix}%
_{i}$as the state of success and the state $%
\begin{pmatrix}
0 \\ 
1%
\end{pmatrix}%
_{i}$ $\left( i=1,2\right) $ as the failure state of $i$ member. The
stationary solution of the model as before has the form Eq. (\ref{k4}), but
now we are interesting in the another question namely :how united group will
be in such process?. Explain what we are keeping in mind. Assume first that
parameters of the model are connected by the relation: $c=a+b$. It is easy
to see in this case that stationary density matrix Eq. (\ref{k4}) can be
represented in the form of direct production of two matrixes: :$\rho \left(
N_{1},N_{2}\right) =\frac{1}{\left( a+b\right) ^{2}}\cdot 
\begin{pmatrix}
ab & 0 & 0 & 0 \\ 
0 & b^{2} & 0 & 0 \\ 
0 & 0 & a^{2} & 0 \\ 
0 & 0 & 0 & ab%
\end{pmatrix}%
=\rho _{1}\otimes \rho _{2}$, where $\rho _{1}=\frac{1}{\left( a+b\right) }%
\begin{pmatrix}
b & 0 \\ 
0 & a%
\end{pmatrix}%
$ and $\rho _{2}=\frac{1}{a+b}%
\begin{pmatrix}
a & 0 \\ 
0 & b%
\end{pmatrix}%
.$ It is natural to interpret such decomposition as the desintegration of
the group. Our main task now is to introduce a quantity which could measure
the unity of the group, i.e. how far is the group from state of
desintegration. For this purpose we will use the analogy of this problem
with a similar problem in quantum theory of composite systems. Remind that
in theory of quantum entanglement to measure how far is given pure state of
composite system from factorized one it is convinient to use the quantity
which called concurrence \cite{7b}. In \ particular if we are interesting
only in two qubit pure states which can be represented in the next vector
form \ $\left\vert \Psi \right\rangle \equiv \left\vert 
\begin{array}{cccc}
z_{1}, & z_{2}, & z_{3}, & z_{4}%
\end{array}%
\right\rangle $ (with normalization condition $\left\vert z_{1}\right\vert
^{2}+\left\vert z_{2}\right\vert ^{2}+\left\vert z_{3}\right\vert
^{2}+\left\vert z_{4}\right\vert ^{2}=1$) the concurrence ( in our case we
prefer to call corresponding quantity as "unity ") can be defined as $%
U=2\left\vert z_{1}z_{4}-z_{2}z_{3}\right\vert $. It is easy to prove that
for all two qubit states $\ 0\leq U\leq 1$. Besides $U=0$ for factorized
states and $U=1$ for maximally entangled (for example four Bell's states).
In our problem however all diagonal elements of density matrix are real
moreover they are positive.Therefore there are no explicit analogs of Bell's
states for our system of interest which is of course classical.Nevertheless
the concept of group unity (i.e. classical analog of concurrence) is still
very useful. We are interested here how group unity depends on parameters of
the model.

According to definition, group unity $U=2\left\vert \frac{a^{2}b^{2}c\left(
a+b-c\right) }{\left[ c\left( a^{2}+b^{2}+ab\right) +ab\left( a+b\right) %
\right] ^{2}}\right\vert $. We consider coefficients $a,b$ in this
expression as fixed parameters and $\ c$ as control parameter and we are
interested in what is the effect of competition on group unity.We can find
the optimal value of $c$ from the condition of maximum $U$: $\frac{\partial U%
}{\partial c}=0$ which implies $c_{op}=\frac{ab\left( a+b\right) }{%
a^{2}+b^{2}+3ab}$. Substituting this value $c_{op\text{ }}$in $U$ we obtain
that $U_{\max }=\frac{ab}{2\left( a+b\right) ^{2}}$. But \ it should be note
that when $c$ tends to infinity the corresponding value of $U$
asymptotically tends to $U_{\infty }=\frac{2a^{2}b^{2}}{\left(
a^{2}+b^{2}+ab\right) ^{2}}$. \ Therefore we must compare two values : $%
U_{\max \text{ \ }}$and $U_{\infty }.$

It is easy to see that this problem reduced to the evaluation of the
expression:\ $F\left( a,b\right) =\left( a^{2}+b^{2}+ab\right)
^{2}-4ab\left( a+b\right) ^{2}$. Condition$F\left( a,b\right) \geq 0$
implies that $U_{\max }\geq U_{\infty }$ and vice versa. Let $b\equiv ta$,
then $F\left( a,b\right) \equiv a^{2}f\left( t\right) =a^{2}\left[ \left(
1+t+t^{2}\right) ^{2}-4\left( 1+t\right) ^{2}\right] $. The equation $%
f\left( t\right) =0$ has two positive roots $t_{1,}t_{2}$ $\left( \text{%
connected by relation }t_{1}t_{2}=1\right) $ which can be found from the
solution of quadratic equation :$t+\frac{1}{t}=1+2\sqrt{2}$. Rough numerical
values of roots are: $t_{1}\approx 0,3,$ $t_{2}\approx 3,3$. and thus we
obtain that $F\left( a,b\right) \leq 0$ in interval $t_{1}\leq t\leq t_{2}$).

It is interesting to note that starting from very simple virtually toy model
we nevertheless come to the conclusion \ which was not obvious in advance.
It turns out that when abilities or efforts of two competing members to
achieve some goal differ not very essentially \ strong competition can
increase the unity of the group. In opposite case when the difference is
significant it is necessary to establish the optimal level of competition to
get the maximal unity of the group.

Let us sum up the results of our study. Starting from ideas of PFT and QTOS
we established the connection between these two theories that seemed far
from each other and proposed the consistent approach for describing simple
phychological phenomena relating both to individuals and groups of
individuals.In the framework of concrete simple model we have demonstrated
the effectiveness of the method proposed to calculate the probabilities of
behavior alternatives and also to predict some peculiarities of behavior
which can hardly be revealed by other approaches. We express the hope that
further development of the method let one to extend its applicability to
more broad sphere of interesting psychological phenomena.

\end{document}